\documentclass[PUB, atlasdraft=false, shownote=true, UKenglish, texmf, orcidlogo]{atlasdoc}
\usepackage{atlaspackage}
\usepackage{threeparttable}
\usepackage{atlasbiblatex}
\usepackage{atlascontribute}

\usepackage{atlasphysics}

\usepackage{booktabs}
\usepackage{multirow}

\graphicspath{{logos/}{figs/}}

\AtlasTitle{Deployment of a Graph-Based Track Seeder for the ATLAS Phase-II Tracking Upgrade}

\AtlasVersion{1.0}

\AtlasAbstract{%
The ATLAS experiment is undergoing significant detector, computing and software upgrades for the High-Luminosity LHC (HL-LHC). The ongoing software modernisation is essential to leverage future processing technologies and ensure maintainability throughout the operation of the experiment, while meeting the tracking performance requirements for physics and reducing the compute needs for event reconstruction. To meet these goals, the ATLAS Collaboration is adopting the ACTS (A Common Tracking Software) toolkit, an experiment-independent software for charged-particle track reconstruction. Using ACTS in the ATLAS reconstruction software at the HL-LHC facilitates the development and integration of new algorithms while promoting thread safety, high performance, and code maintainability. This document describes the development and recent deployment of a Graph-Based Track Seeder (GBTS), its underlying methodology, and its expected tracking performance at the HL-LHC.
}

\author{The ATLAS Collaboration}

\AtlasRefCode{ATL-PHYS-PUB-2026-030}

\AtlasNote{ATL-PHYS-PUB-2026-030}

\hypersetup{pdftitle={ATLAS document},pdfauthor={The ATLAS Collaboration}}

\begin{document}

\maketitle

\section{Introduction}
\label{sec:intro}

In recent years, the ATLAS Collaboration has worked on an extensive detector~\cite{ATLAS-LoI-PhaseII,ATLAS-TDR-PhaseII} and software upgrade programme~\cite{ATLAS-CDR-Computing,ATLAS:2022SoftwareRoadmap} to cope with the high instantaneous luminosity and the number of simultaneous proton-proton interactions per bunch crossing ($\mu$) expected at the High-Luminosity LHC (HL-LHC)~\cite{ZurbanoFernandez:2020cco}, which are expected to be significantly higher than in LHC Run\,3. An entirely new all-silicon Inner Tracker (ITk)~\cite{ATLAS-TDR-30,ATLAS-TDR-25} will be installed during Long Shutdown 3, which started in 2026. It is composed of Pixel and Strip subsystems extending the tracking coverage to $|\eta|~<~4.0$ from the Run\,1--3 coverage of $|\eta|<2.5$. Together with the detector upgrade, a software modernisation is also taking place. One of the areas of major activity is track reconstruction, where components from the ACTS (A Common Tracking Software) toolkit~\cite{Ai:2021qau} are being integrated and deployed in the ATLAS reconstruction software~\cite{SOFT-2022-02,ATL-PHYS-PUB-2024-017}. The main goal is meeting the tracking performance requirements for physics while reducing the compute requirements for event reconstruction and improving code maintainability.

The reconstruction of tracks from charged particles traversing the ITk is an essential part of event reconstruction at the ATLAS experiment~\cite{IDTR-2023-01}. This is achieved through a sequence of algorithms. First, a clustering algorithm groups signals from individual channels within the ITk Pixel and Strip subdetectors that are likely to originate from the same particle. These groups of signals are referred to as clusters. The clusters are then converted into three-dimensional space points, which are used by the track seeding algorithm to form track seeds and provide initial estimates of the track parameters. These seeds serve as the initial state for the track-finding algorithm, which is based on a Combinatorial Kalman Filter~\cite{Fruhwirth:1987fm,Mankel:1997dy}. The algorithm extends the seeds into track candidates by incorporating all clusters compatible with the predicted trajectory. In the track reconstruction chain considered so far, the resulting track candidates are subsequently processed by an ambiguity-solving algorithm to resolve cases where tracks share multiple clusters, reject low-quality candidates, and perform the final track fit. This dedicated step is no longer required in the ACTS-based reconstruction chain, which makes use of precise material modelling, accurate magnetic-field maps and refined calibrations during track finding, so that high-precision track parameters are obtained without the separate, computationally demanding refit previously required.

Since Run 1, track seeds produced with seeding algorithms have consisted of triplets of space points from the silicon parts of the Inner Detector. To study the expected tracking performance of ATLAS at the HL-LHC and validate the new track reconstruction algorithms ahead of Run 4 data taking, the triplet seeding algorithm has been adapted to account for the expected HL-LHC conditions and the specific geometry of the ITk. In the higher pile-up environment, the combinatorial complexity increases substantially, leading to a low seed purity, i.e. a large fraction of seeds that lead to duplicate or fake track candidates. The track finding has to reject these candidates, which increases the compute requirements for track reconstruction. While further developments and optimisation were carried out in the most recent years, the seeding algorithm remained among the most computationally expensive components of the tracking chain, at the same level as the track-finding algorithm~\cite{IDTR-2022-04}.

The Graph-Based Track Seeder (GBTS) is a seeding algorithm designed for high pile-up conditions such as those expected during ATLAS Run 4 data taking. This algorithm was originally developed as \texttt{fastrack} as part of the TrackML challenge~\cite{Amrouche:2021nbs}, and then implemented in the ATLAS software. Subsequently, it was reimplemented directly in ACTS to ensure seamless integration with the toolkit and consistency with the ATLAS software's broader adoption of ACTS for track reconstruction.

The algorithm addresses the issue of high combinatorics through two complementary approaches. First, it reduces the seeding runtime by applying computationally inexpensive selection criteria, exploiting the assumption that prompt tracks originate from the beam spot. Second, it constructs a graph connecting compatible space points, enabling the production of high-purity seeds. Through these two improvements, GBTS reduces the computing time substantially without compromising track reconstruction efficiency and quality and has been deployed as part of the ATLAS offline track reconstruction.

The note is structured as follows: \cref{sec:GBTS} provides a description of the GBTS seeding strategy for prompt tracks, its implementation in ACTS and deployment in the ATLAS reconstruction software; \cref{sec:result} summarises the obtained track reconstruction performance using GBTS; finally \cref{sec:conclusion} concludes the note.

\section{The Graph-Based Track Seeder algorithm}
\label{sec:GBTS}

The GBTS algorithm operates on nodes and edges aiming at constructing chains of nodes likely corresponding to true particle tracks. Nodes are objects created from space points: they hold the positions of the original clusters in the ATLAS global reference frame coordinates\footnote{ATLAS uses a right-handed coordinate system with its origin at the nominal interaction point (IP) in the centre of the detector and the $z$-axis along the beampipe. The $x$-axis points from the IP to the centre of the LHC ring, and the $y$-axis points upwards. Polar coordinates $(r,\phi)$ are used in the transverse plane, $\phi$ being the azimuthal angle around the $z$-axis. The pseudorapidity is defined in terms of the polar angle $\theta$ as $\eta = -\ln \tan(\theta/2)$ and is equal to the rapidity $ y = \frac{1}{2} \ln \left( \frac{E + p_z}{E - p_z} \right) $ in the relativistic limit.}, their $\eta$-width\footnote{The size of the cluster in the direction of $\eta$, which corresponds to the $z$-axis in the barrel region and to the transverse direction $r$ in the endcap region.} and a layer index. A layer in GBTS is a grouping of modules corresponding to a geometrical structure in the detector geometry. Edges are connections between two compatible nodes. Nodes and edges are used in the different stages of the GBTS algorithm, corresponding to graph construction, seed finding and seed filtering.

\subsection{Graph construction}
\label{subseubsection: Graph Construction}

\begin{figure}
\centering
\includegraphics[width=0.9\linewidth]{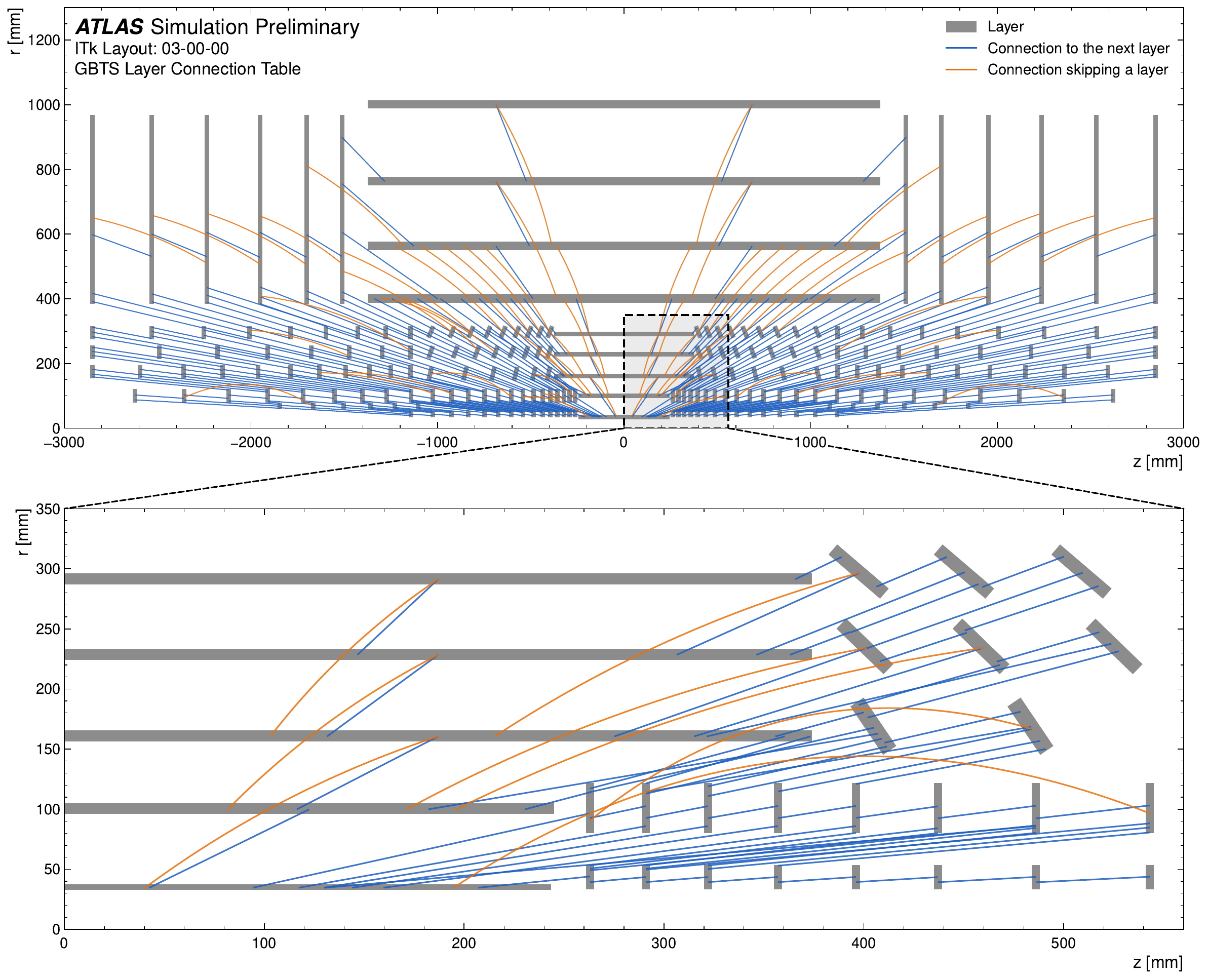}
\caption{The layer connection table for GBTS of the ITk detector, with a subset of the ITk Pixel subdetector highlighted. The geometrical layers, which correspond to the grouping of detector modules, are represented in grey. The direct connections between layers are illustrated as blue lines, while the connections that skip a layer are shown with orange lines. Only nodes on valid layer pairs are checked for compatibility to form an edge.}
\label{fig: Layer linking scheme}
\end{figure}

The GBTS algorithm aims to create high-purity seeds. To achieve this, connections are first restricted to pairs of space points belonging to physically compatible detector layers, thereby preventing the formation of invalid edges. A layer connection table is created using Monte Carlo simulated samples of 100\,000 single-muon events with $p_{\mathrm{T}}~=~1$ GeV. This momentum gives the largest multiple scattering expected for valid tracks, so the table also covers all layer pairs of tracks at higher $p_{\mathrm{T}}$. \cref{fig: Layer linking scheme} shows this table for the ITk geometry, with a zoom on the inner Pixel region. The algorithm described in this note uses only Pixel space points and hence restricts the layer connection table to the Pixel subdetector.

The width of the cluster in the $\eta$-direction provides an additional criterion for edge selection. Under the assumption that particles originate within a given $z$-window around the interaction point, the cluster width defines a valid $\eta$-interval for each node. Only edges connecting nodes with overlapping $\eta$-intervals are retained.

At this point, for each layer, nodes are placed into fixed-width $\eta$ bins and sorted in ascending $\phi$ order. The algorithm then makes use of the valid layer connections and creates pairs of bins that are compatible with a straight line from the interaction point.

\begin{figure}
\centering
\includegraphics[width=0.7\textwidth]{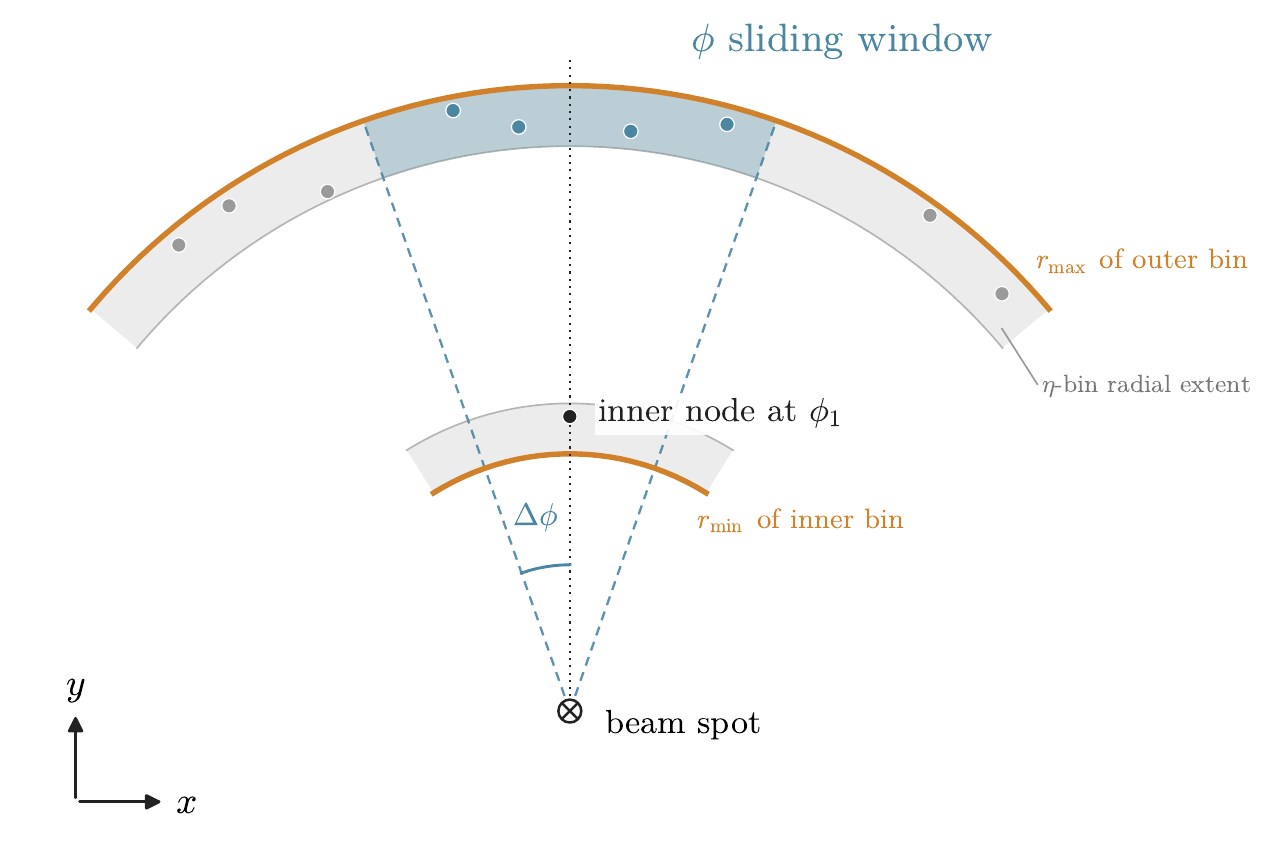}
\caption{Illustration of the sliding $\phi$ window projected from an inner node to reduce the region for valid connections to outer nodes.}
\label{fig:phi_window}
\end{figure}

A window in $\phi$ is defined by the maximum curvature (i.e. minimum $p_{\mathrm{T}}$) allowed by a track traversing from the inner to outer $\eta$ bin. This window is then centred at each inner node such that only a small subset of connections to outer nodes need to be checked. Edges are constructed from nodes within the $\phi$ window, imposing additional constraints based on the allowed range for track parameters (mainly transverse and longitudinal impact parameters and transverse momentum) and  cluster $\eta$-widths. \cref{fig:phi_window} illustrates a $\phi$ window derived from a node on the innermost layer. This process continues for all $\phi$ windows and $\eta$-bin pairs, between connected layers. With all the edges created, any that share common nodes are linked into connected components. Those with fewer nodes than a configured minimum are discarded.

\subsection{Seed finding and filtering}
\label{subsubsection: Seed Finding and Filtring}

The seed extraction from the graph follows two steps.

\begin{figure}
\centering
\includegraphics[width=1\linewidth]{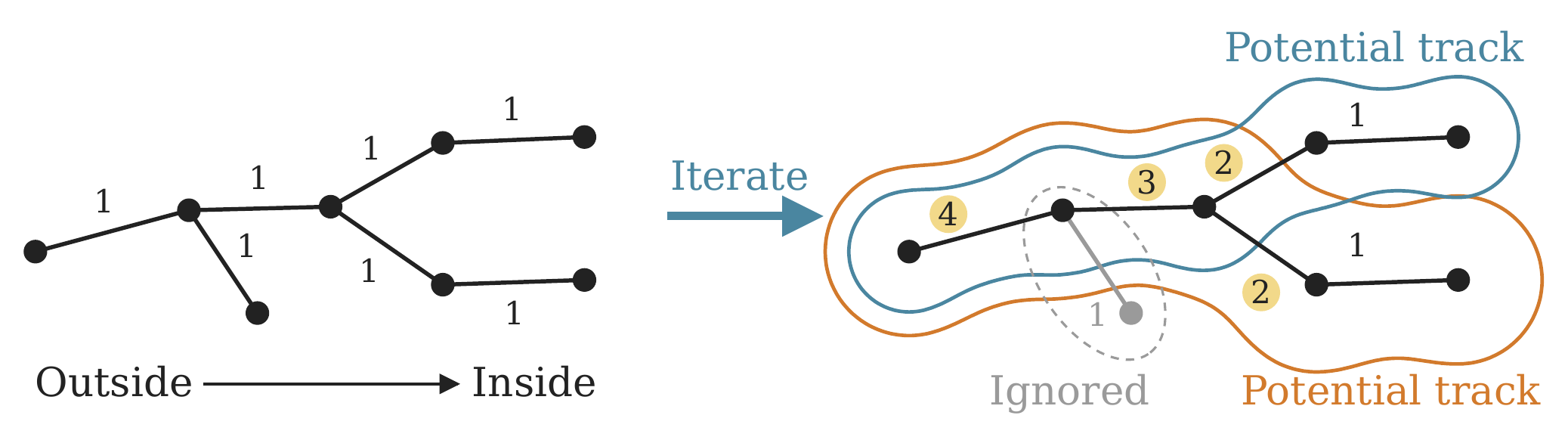}
\caption{The Cellular Automaton (CA) updates the levels of the edges by 1 for each iteration if a neighbouring edge closer to the interaction point has the same level. After three iterations, the levels are updated to the values marked with the yellow circles. This is used to reject branches that are unlikely to be produced by a true particle trajectory.}
\label{fig: CCA}
\end{figure}

First, a Cellular Automaton (CA)-based seed finding~\cite{CATS-HERA-B} is used to extract longer chains of connected edges from the graph. This works as follows: the CA starts by assigning an integer value of 1, called the level, to every edge in the graph. Then, starting from the outermost edges, it looks at the connected neighbouring edges towards the interaction point and updates its edge level. If at least one of those neighbours has the same level, the level of the outer edge is increased by 1. This update is repeated until no more updates occur. The level of an edge corresponds to the number of edges in the longest chain that extends inward from that edge, including the edge itself. The outermost edges are used as the starting points to construct seed candidates by connecting only edges that have a level of exactly 1 below the previous one. This approach is very effective at removing dead-end branches from the seed candidates. An example of the edge level update procedure and edge tracing is illustrated in \cref{fig: CCA}.

The seed candidates produced by the CA are then passed to a Kalman Filter (KF)~\cite{Fruhwirth:1987fm}. The KF follows the seed candidates and scores each node according to its compatibility with an approximate helix trajectory. Early rejection is performed on seed candidates for which the $\chi^2$ value of at least one node exceeds a predefined threshold. The remaining seed candidates are scored based on the number of nodes in the seed and their $\chi^{2}$ value. Seed candidates for which the majority of nodes are shared with a higher-scoring candidate are removed. Finally, a seed splitting step is used for seed candidates with four or five space points within $|\eta| < 0.6$. This samples three space points from these chains and checks their compatibility; if there is a discrepancy larger than a set value, the seed is split. In the case of candidates with four space points, this can result in seeds of three space points. The number of seeds that go through this process is small in general but can help with regaining a small amount of efficiency. The final collection of seed candidates is used to estimate the initial track parameters for the track-finding algorithm. This estimate relies on the first, middle and last space points of each seed, thereby using the full radial extent of the seeds.

\section{Results}
\label{sec:result}

All components of the GBTS algorithm have been implemented in ACTS and then deployed in the ATLAS ACTS-based track reconstruction chain~\cite{ATL-PHYS-PUB-2024-017}. This section summarises the expected seeding and tracking performance and compares it with that obtained using the Run 3-based tracking chain, hereafter referred to as Legacy~\cite{IDTR-2023-01}. The results are obtained using a $t\bar{t}$ event sample at $\sqrt{s}=14$~TeV with pile-up ranging from 190 to 210, labelled as $\langle\mu\rangle=200$. The number of pile-up interactions is randomly sampled from a Poisson distribution with a mean corresponding to the average pile-up. The production of the \ttbar sample is modelled using the \POWHEGBOX[v2]~\cite{Frixione:2007nw,Nason:2004rx,Frixione:2007vw,Alioli:2010xd} generator at NLO with the \NNPDF[3.0nlo]~\cite{Ball:2014uwa} PDF set and the $h_{\mathrm{damp}}$ parameter\footnote{The \hdamp parameter is a resummation damping factor and one of the parameters that controls the matching of \POWHEG matrix elements to the parton shower and thus effectively regulates the high-\pT radiation against which the $t\bar{t}$ system recoils.} fixed to $1.5m_{\mathrm{top}}$~\cite{ATL-PHYS-PUB-2016-020}, with $m_{\mathrm{top}}$ = 172.5~\GeV used for the top-quark mass. The events were interfaced to \PYTHIA[8.307]~\cite{Sjostrand:2014zea} to model the parton shower, hadronisation, and underlying event, with parameters following the A14 tune~\cite{ATL-PHYS-PUB-2014-021} and using the \NNPDF[2.3lo] set of PDFs~\cite{Ball:2012cx}. The decays of bottom and charm hadrons were performed by \EVTGEN[2.1.1]~\cite{Lange:2001uf}. Only \ttbar events with a single leptonic $W$-boson decay have been considered.

\begin{figure}[htbp]
\centering
\subfloat[\label{fig: seed eff vs eta}]{%
\centering
\includegraphics[width=0.47\textwidth]{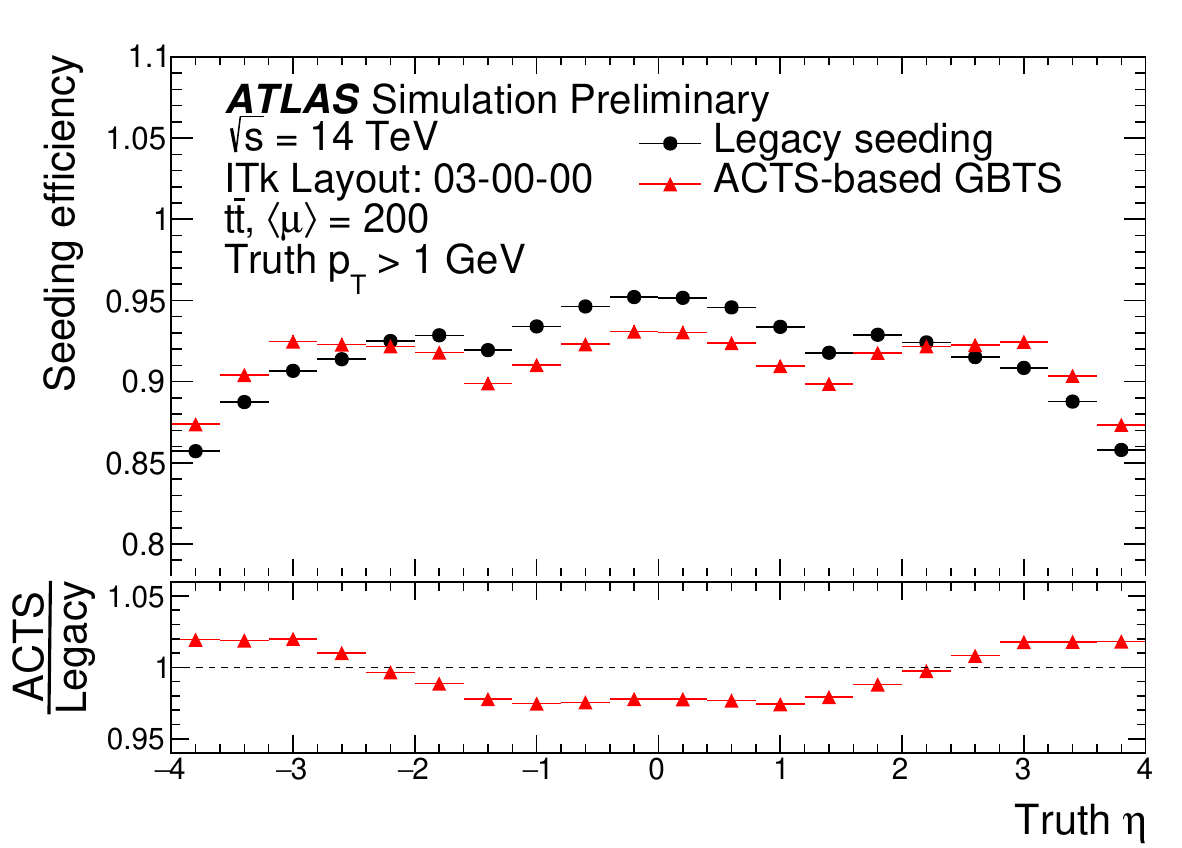}
}
\hfill
\subfloat[\label{fig: seed eff vs pT}]{%
\centering
\includegraphics[width=0.47\textwidth]{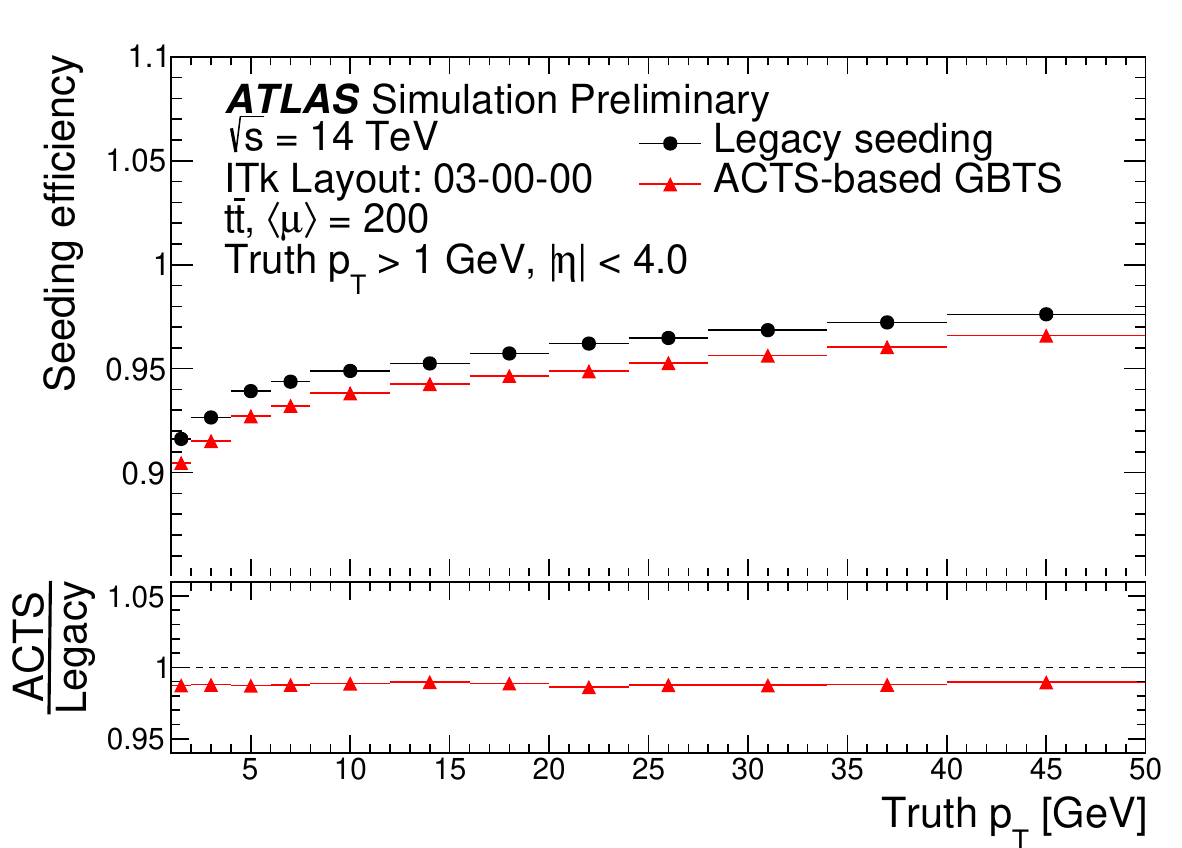}
}
\caption{Comparison of the expected seeding efficiencies as a function of $\eta$ (a) and \pT (b) for $t\bar{t}$ events at $\langle\mu\rangle = 200$ for hard-scatter particles with $p_\mathrm{T} > 1$~GeV for the GBTS algorithm and the Legacy triplet seeding algorithm.}
\end{figure}

\begin{figure}[htbp]
\begin{minipage}[t]{0.47\textwidth}
\centering
\includegraphics[width=\textwidth]{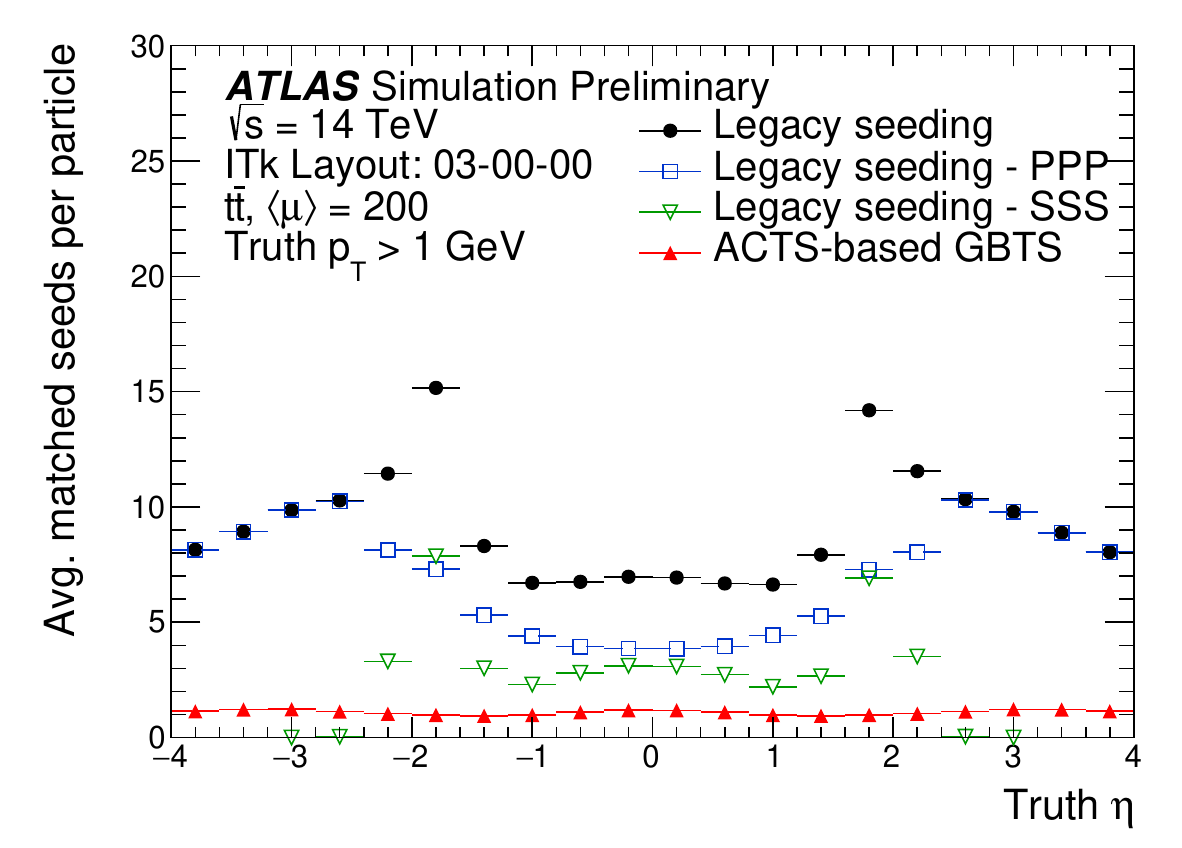}
\caption{Comparison of the average number of matched seeds per particle as a function of $\eta$ for $t\bar{t}$ events at $\langle\mu\rangle = 200$ for hard-scatter particles with $p_\mathrm{T} > 1$~GeV for the GBTS algorithm and the Legacy triplet seeding algorithm. For the latter, the numbers are shown separately for Pixel-only (PPP), Strip-only (SSS) and both combined.}
\label{fig: seed duplicate rate vs eta}
\end{minipage}
\hfill
\begin{minipage}[t]{0.47\textwidth}
\centering
\includegraphics[width=\textwidth]{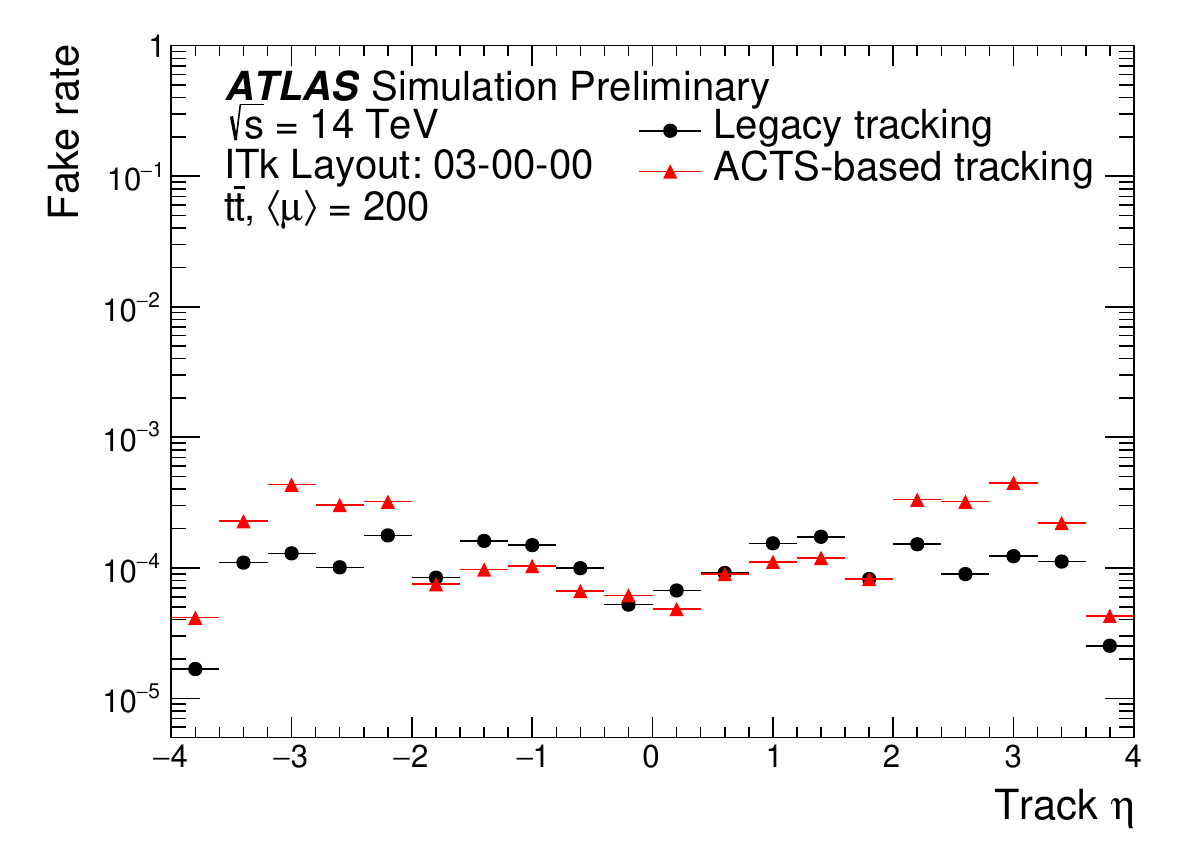}
\caption{The rate of fake tracks as a function of $\eta$ for $t\bar{t}$ events at $\langle\mu\rangle = 200$ for the ACTS-based and Legacy tracking chains.}
\label{fig: track fake rate vs eta}
\end{minipage}
\end{figure}

\begin{figure}[htbp]
\centering
\subfloat[\label{fig: track eff vs eta}]{%
\centering
\includegraphics[width=0.47\textwidth]{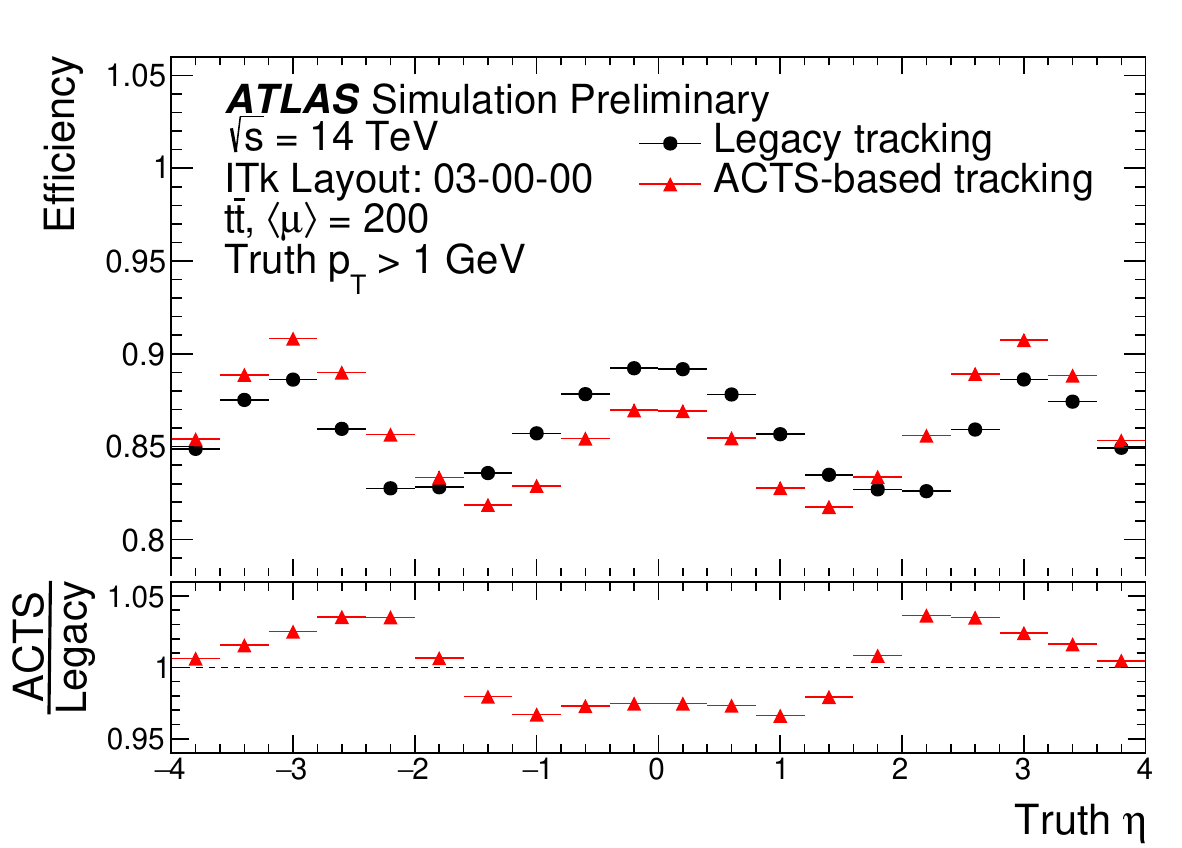}
}
\hfill
\subfloat[\label{fig: track eff vs pT}]{%
\centering
\includegraphics[width=0.47\textwidth]{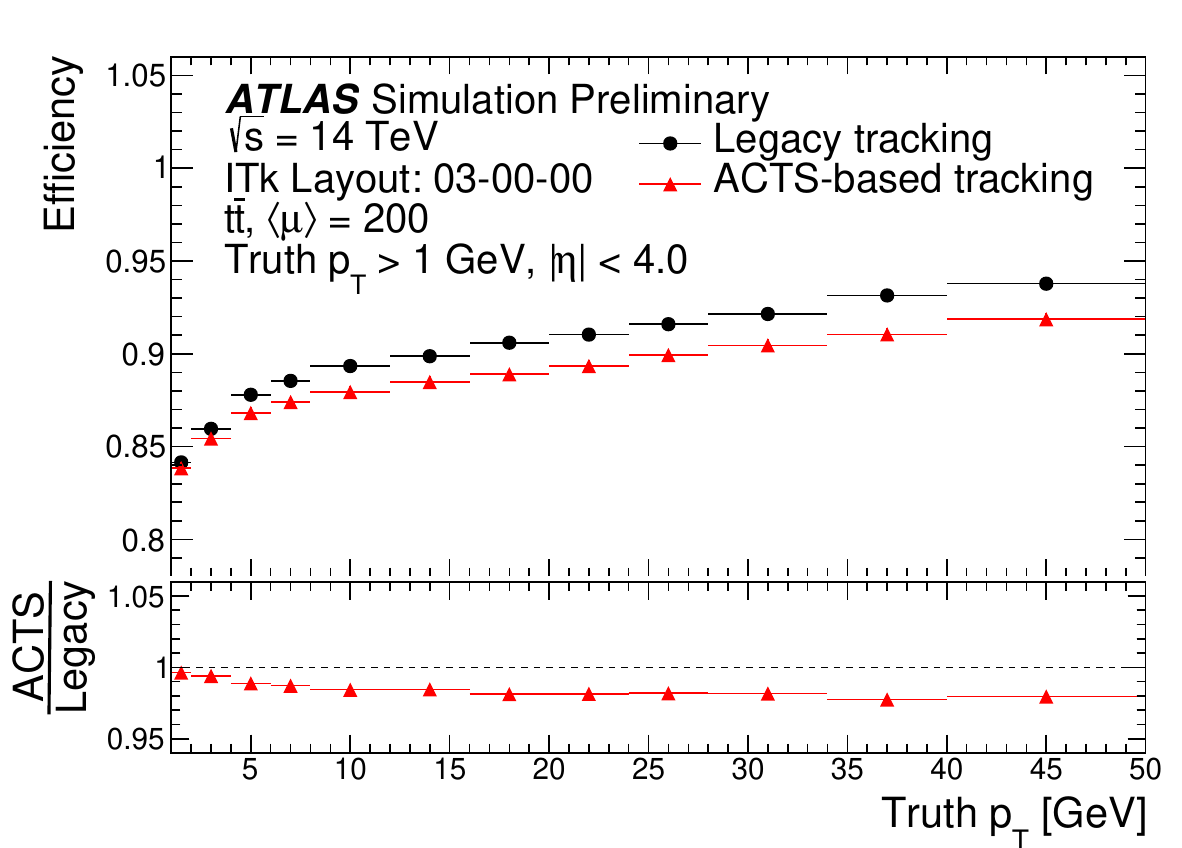}
}
\caption{Comparison of the expected tracking efficiencies as a function of $\eta$ (a) and \pT (b) for $t\bar{t}$ events at $\langle\mu\rangle = 200$ for hard-scatter particles with $p_\mathrm{T} > 1$~GeV for the ACTS-based and Legacy tracking chains.}
\end{figure}

The performance of the seeding algorithm is assessed using different figures of merit that rely on associating seeds with stable charged particles produced by the Monte Carlo simulation. A seed is considered matched to a particle if more than half of its measurements originate from this particle, and the fraction of particles matched to at least one seed is defined as the seeding efficiency. Only particles with $p_\mathrm{T} > 1$~GeV within the detector acceptance ($|\eta| < 4.0$), stemming from the hard-scatter interaction, are considered.

\cref{fig: seed eff vs eta,fig: seed eff vs pT} show the seeding efficiency of GBTS in the ACTS-based track reconstruction chain and of the Legacy triplet seeder. While the Legacy seeding algorithm produces triplet seeds both in the ITk Pixel and Strip subdetectors, the GBTS algorithm is currently configured to find seeds only using Pixel space points. In both configurations, Pixel seed candidates are required to contain at least four space points. The Legacy seeding algorithm achieves this by requiring a fourth space point to confirm the trajectory of an initial triplet seed. The Strip seeds of the Legacy configuration help to recover seeding efficiency within $|\eta|<1.5$ when too few measurements are available in the Pixel detector to form a valid seed. This can occur, for example, when a particle traverses a gap between Pixel modules or when the charge deposited in the active detector area is insufficient to activate a readout channel. GBTS allows for a higher seeding efficiency at $|\eta|>2.5$, as a result of looser requirements in the edge construction compared to the ones for triplet formation in the Legacy seeder. As a function of \pT, the GBTS seeding efficiency is about 1\% to 1.5\% below that of the Legacy seeder over the entire range.

The average number of matched seeds per particle obtained with GBTS and the Legacy triplet seeding algorithm is shown in \cref{fig: seed duplicate rate vs eta}, highlighting the difference between the two strategies: while the Legacy seeding algorithm by construction produces multiple triplet seeds per particle (from both Pixel and Strip subdetectors), GBTS allows for one long seed per track, where the connected space points are likely to be associated with the same particle. Reconstructing multiple seeds per particle can be beneficial to robustness against detector inefficiencies and misalignment. The impact of these effects using the GBTS algorithm is under evaluation.

Analogously to the seeding efficiency, the tracking efficiency is defined as the fraction of charged particles associated with a reconstructed track. The differences in seeding efficiency are also reflected in the tracking efficiency, as shown in \cref{fig: track eff vs eta,fig: track eff vs pT}.

\cref{fig: track fake rate vs eta} shows the rate of fake tracks as a function of $\eta$. The fake rate for the ACTS-based tracking chain is compatible with the one of the Legacy tracking, and well under control; the fake rate obtained with GBTS is larger than the Legacy one for $|\eta|>2$, corresponding to the region of higher seeding and tracking efficiency.

\begin{figure}[htbp]
\centering
\subfloat[$d_0$ resolution\label{fig: track d0 vs eta}]{%
\includegraphics[width=0.47\textwidth]{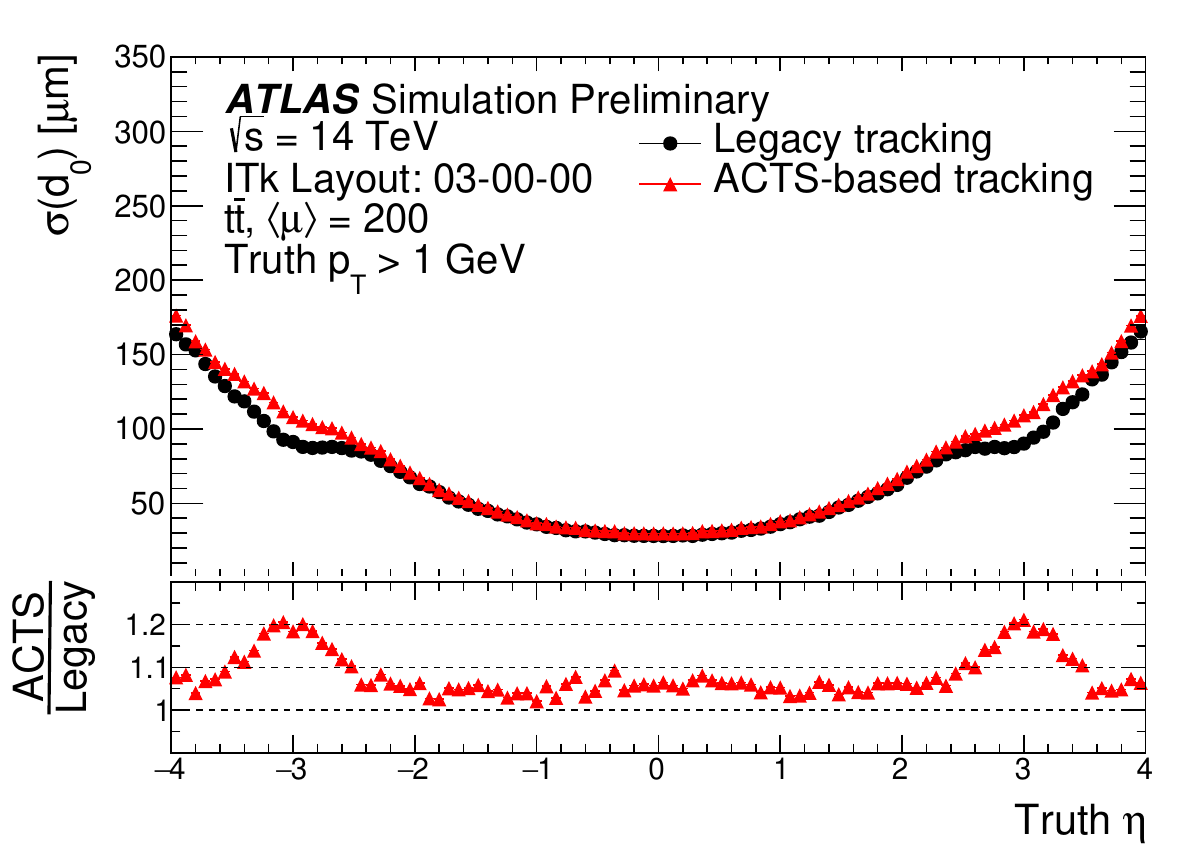}}
\hfill
\subfloat[$z_0$ resolution\label{fig: track z0 vs eta}]{%
\includegraphics[width=0.47\textwidth]{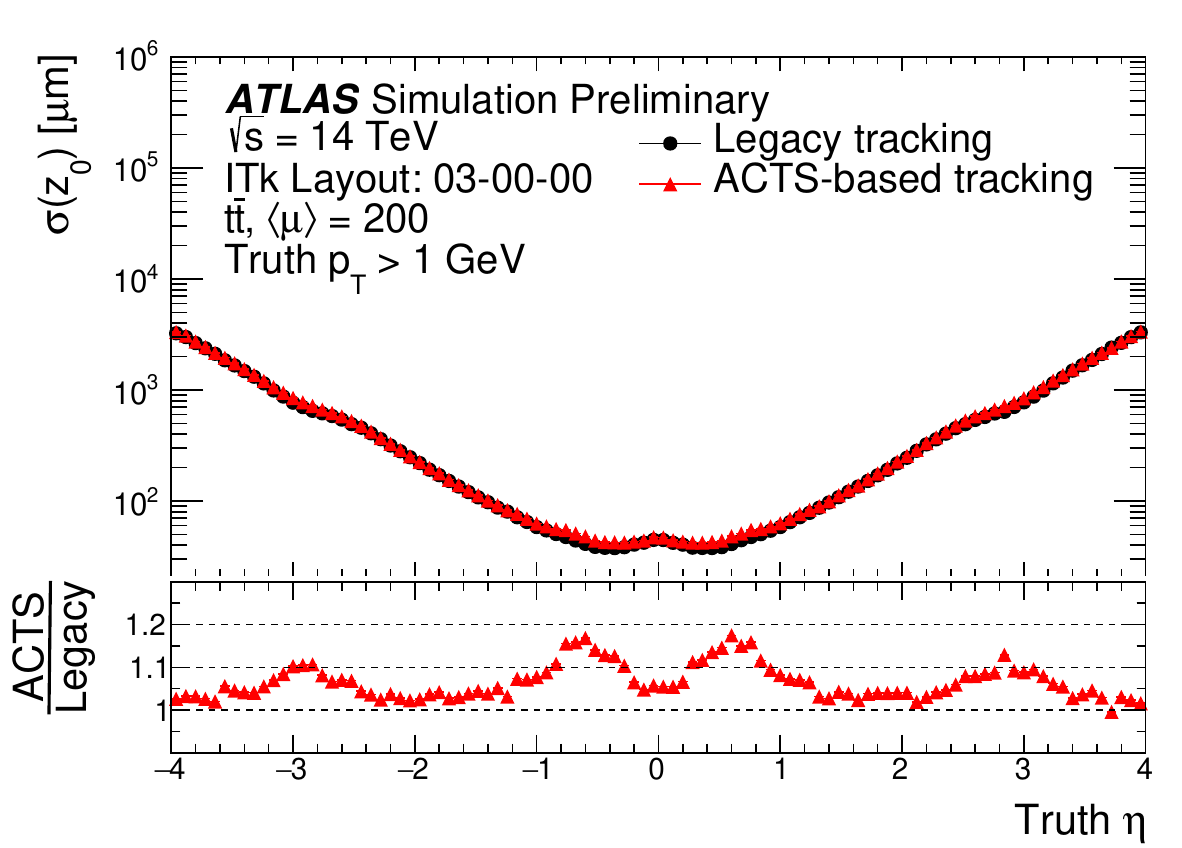}}
\caption{Comparison of the transverse $d_0$ (a) and longitudinal $z_0$ (b) impact parameter resolutions as a function of $\eta$ for $t\bar{t}$ events at $\langle\mu\rangle = 200$ for hard-scatter particles with $p_\mathrm{T} > 1$~GeV for the ACTS-based and Legacy tracking chains.}
\label{fig: track impact parameter vs eta}
\end{figure}

The resolution on track parameters is estimated from reconstructed tracks associated with charged particles. \cref{fig: track d0 vs eta,fig: track z0 vs eta} show the expected resolution of the transverse ($d_0$) and longitudinal ($z_0$) impact parameters for tracks in $t\bar{t}$ events at $\langle\mu\rangle=200$. The ACTS-based tracking chain exhibits a slightly worse resolution compared to the Legacy chain. The ACTS-based tracking chain is still under active development. At the time of writing, several important features aimed at improving impact parameter resolution were still being developed, including optimal calibration of measurement positions and covariances. Further investigations are ongoing to improve the parameter estimation for long seeds, which are more common at $|\eta|>2$. In this region, the seed parameters, estimated using the first, middle and last space points, are subject to biases arising from larger multiple-scattering contributions and the non-zero radial component of the magnetic field towards the endcaps.

Finally, Table \ref{tab:timings} summarises the timing results for the Legacy tracking chain and for two ACTS-based configurations that differ only in the seeding algorithm. For the Legacy chain, the time is split into the combined seeding and track finding step, and the ambiguity solving step. These are defined by an ACTS implementation of a Legacy-like triplet seeder~\cite{Allaire:2943308} and the GBTS algorithm described in this note. Both ACTS-based workflows achieve a substantial speed-up compared with the Legacy reconstruction chain. The ACTS-based strategy requires no dedicated ambiguity-solving step, because the track finding already resolves cluster–track ambiguities and suppresses duplicates. It also requires no dedicated track-fitting step, historically part of ambiguity solving, because the track finding already outputs fitted tracks. In addition, GBTS reduces the seeding time by approximately a factor of five compared with the Legacy-like triplet seeder, while maintaining physics performance close to that of the Legacy configuration, as demonstrated in the previous sections. In total, the ACTS-based chain with GBTS is faster than the Legacy chain by a factor of about 13, and faster than the ACTS-based chain with the triplet seeder by a factor of about 2. Furthermore, owing to the increased seed purity and reduced amount of seeds, the track finding time decreases slightly, within the uncertainty of the measurement.

\begin{table}[htbp]
\centering
\caption{Per-event timings for seeding, track finding and ambiguity solving. Measured on $t\bar{t}$ events with $\langle\mu\rangle = 200$, single-threaded, on an Intel Xeon Gold 6326 at 2.90\,GHz. The uncertainty on the measurements is 5\%, corresponding to the observed timing variations on the machine.}
\label{tab:timings}
\begin{tabular}{llcccc}
\toprule
\multirow{2}{*}{Workflow} & \multirow{2}{*}{Seed method} & \multicolumn{4}{c}{Time per event [ms]} \\
\cmidrule(l){3-6}
& & Seeding & Track finding & Ambiguity solving & Sum \\
\midrule
Legacy & Triplet Pixel + Strip & \multicolumn{2}{c}{5100} & 3500 & 8600 \\
\midrule
\multirow{2}{*}{ACTS-based} & Triplet Pixel & 860 & 530 & -- & 1390 \\
& GBTS Pixel    & 180 & 490 & -- &  670 \\
\bottomrule
\end{tabular}
\end{table}

\section{Conclusion}
\label{sec:conclusion}

This note summarises the deployment of the Graph-Based Track Seeder (GBTS) for ATLAS track reconstruction at the HL-LHC. It describes the algorithm implemented within the ACTS toolkit and demonstrates its performance following integration into the ATLAS track reconstruction software. Both tracking and computational performance are presented, demonstrating a substantial reduction in computing time while maintaining high tracking efficiency, good track-parameter resolutions and a low fake rate. Since GBTS significantly improves computational performance while maintaining broadly comparable physics performance to the ACTS-based implementation of the Legacy-like triplet seeder, it has been selected as the default seeding algorithm for ATLAS offline track reconstruction. The improved computational performance provides greater flexibility for further optimisation of the ACTS-based tracking chain. Ongoing work focuses on addressing the remaining differences with respect to the Legacy tracking chain, in particular by reducing the fake rate, improving the impact parameter resolutions and better understanding the observed performance differences.
Overall, GBTS has demonstrated strong computational performance while achieving broadly comparable physics performance, making it a promising candidate for track reconstruction in the HL-LHC era.

\printbibliography

\end{document}